\documentclass[11pt,a4paper]{article}
\usepackage[utf8]{inputenc}
\usepackage[left=1in,right=1in,top=1in,bottom=1in]{geometry}
\usepackage{amsmath}
\usepackage{amsthm}
\usepackage{amsfonts}
\usepackage{amssymb}
\usepackage{cite}
\usepackage{mathtools}
\usepackage{algorithm}
\usepackage[noend]{algpseudocode}
\usepackage{float}
\usepackage{multicol}
\usepackage{graphicx}
\usepackage{authblk}
\usepackage{verbatim}
\usepackage{ctable}
\usepackage{authblk}
\usepackage{makecell}
\usepackage{caption}
\usepackage{fancyhdr}
\usepackage{color}
\usepackage{hyperref}
\usepackage[section]{placeins}
\usepackage{lineno}
\begin{document}
\title{Segmental Lennard-Jones Interactions for Semi-flexible Polymer Networks}
\author[1]{Carlos Floyd}

\author[2]{Aravind Chandresekaran}

\author[1]{Haoran Ni}

\author[3]{Qin Ni}

\author[1,2,4]{Garegin A. Papoian}

\affil[1]{
	Biophysics Program, University of Maryland, College Park, MD 20742 USA
}
\affil[2]{Department of Chemistry and Biochemistry, University of Maryland, College Park, MD 20742 USA}
\affil[3]{Department of Chemical and Biomolecular Engineering, University of Maryland, College Park, MD 20742 USA}
\affil[4]{
	Institute for Physical Science and Technology, University of Maryland, College Park, MD 20742 USA}

\date{\today}

\maketitle

\begin{abstract}
	
Simulating soft matter systems such as the cytoskeleton can enable deep understanding of experimentally observed phenomena. One challenge of modeling such systems is realistic description of the steric repulsion between nearby polymers.  Previous models of the polymeric excluded volume interaction have the deficit of being non-analytic, being computationally expensive, or allowing polymers to erroneously cross each other.  A recent solution to these issues, implemented in the MEDYAN simulation platform, uses analytical expressions obtained from integrating an interaction kernel along the lengths of two polymer segments to describe their repulsion.  Here, we extend this model by re-deriving it for lower-dimensional geometrical configurations, deriving similar expressions using a steeper interaction kernel, comparing it to other commonly used potentials, and showing how to parameterize these models. We also generalize this new integrated style of potential by introducing a segmental Lennard-Jones potential, which enables modelling both attractive and repulsive interactions in semi-flexible polymer networks.  These results can be further generalized to facilitate the development of effective interaction potentials for other finite elements in simulations of soft-matter systems. 

\end{abstract}

\newpage
\null
\vfill
\begin{center}
\begin{figure}[H]
		\includegraphics[width=16 cm]{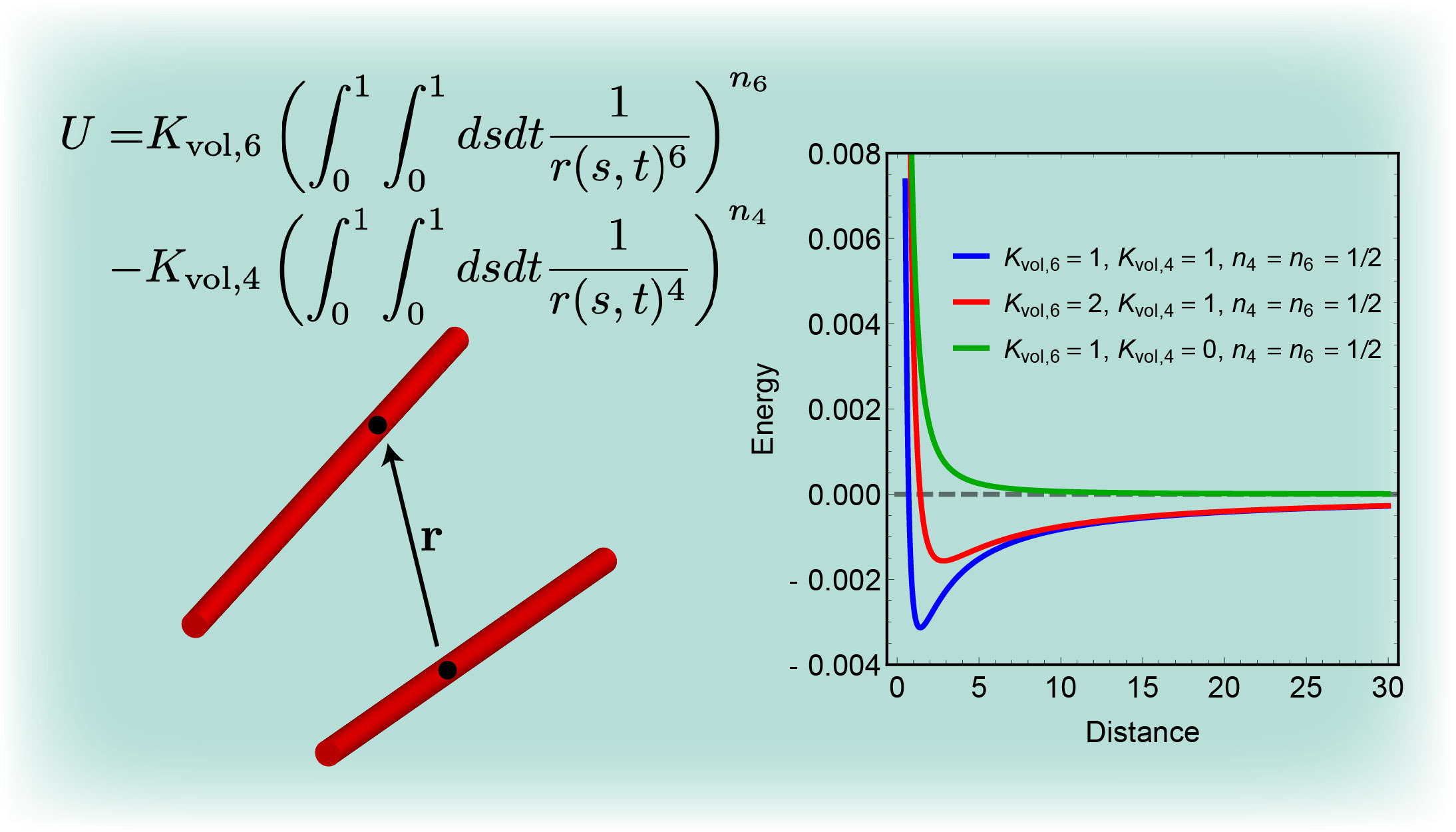}
		\caption*{Graphical Abstract: The segmental Lennard-Jones potential allows for tunable attractive and repulsive interactions between finite cylindrical segments.}

\end{figure}
\end{center}
\vfill
\clearpage

\begin{multicols}{2}
\section{Introduction}

Excluded volume interactions between spatially extended macromolecules play an important role in a wide range of cellular phenomena.  They help to produce mesoscopically ordered structures, which enable the complex functionality exhibited by cells. For example, it has been shown that steric interactions alone can induce alignment of the long biopolymers that comprise the cytoskeleton \cite{lopez2014vitro}.  These interactions have also been implicated in transmitting non-equilibrium fluctuations from one cellular subsystem to another \cite{brangwynne2008nonequilibrium}.  Excluded volume, or steric, interactions are in fact an important physical feature of many soft matter systems, which are often controlled by a complex interplay of steric and entropic effects \cite{doi2013soft, rubinstein2003polymer}.

The computational modeling of such soft matter systems has become an essential tool used in biology, chemistry, and physics \cite{schiller2018mesoscopic, gartner2019modeling}.  In particular, software packages for simulating the cytoskeleton, a complex soft active matter system comprising interlinked biopolymers and molecular motors, have helped provide theoretical understanding of various experimental phenomena \cite{howard2001mechanics, li2020tensile, chandrasekaran2019remarkable, floyd2019quantifying, freedman2019mechanical, belmonte2017theory}.  Reaching timescales of thousands of seconds and length scales of tens of micrometers, packages such as AFiNeS, CytoSim, the model of Kim and coworkers, and MEDYAN allow exploration of fascinating emergent cytoskeletal phenomena while striving to preserve realistic microscopic physics \cite{freedman2017versatile,nedelec2007collective,kim2009computational,popov2016medyan}.  These models typically employ effective, coarse-grained potentials based on ideas from polymer physics.  For example, treating a semi-flexible polymer (for which the typical polymer length is comparable to the persistence length) as a one-dimensional piecewise linear chain, the mechanical strain energy can be straightforwardly decomposed into stretching and bending terms which can be computed using harmonic functions of the linear segments' positions \cite{broedersz2014modeling}.  It is less straightforward, however, to model the potential energy mediating the excluded volume interaction between neighboring polymers.  This is treated in different ways between CytoSim, the model of Kim and coworkers, and MEDYAN, and it is not considered in AFiNeS.  However, accurately modelling repulsion between polymers is essential for realistically simulating important behaviors such as entanglement, reptation, liquid crystal ordering, and entropic depletion forces \cite{rubinstein2003polymer, de1971reptation, gotzelmann1998depletion, chaikin1995principles}.

The primary physical origin of the excluded volume interaction between typical biopolymers such as actin is screened Coulomb repulsion \cite{phillips2012physical, rubinstein2012polyelectrolytes, tadmor2002debye}.  Actin filaments have a relatively high linear charge density ($\sim 0.4 \ e / \AA$), but biological ionic environments have a Debye-H\"uckel screening length ($\sim 1$ nm) the same order of magnitude as the filament radius ($\sim 3.5$ nm) \cite{angelini2006counterions, smith2011electrostatic}.   In specialized tightly packed actin bundles and sarcomeric structures (with inter-filament spacings $\sim 0.3 - 30$ nm) complicated ion distributions are established, but in more common actin cortical networks the screening length is much less than the average inter-filament spacing ($\sim 30 - 150$ nm)  \cite{eghiaian2015structural, smith2011electrostatic, sanders2005structure, janmey1994mechanical}.   For typical actin networks, therefore, a suitable approximation to the interaction of screened, electrically charged polymers is a hard-wall potential; however this is a discontinuous function poorly suited to implementation in dynamical simulations.  Additionally, it is not immediately clear how to define the distance between two linear segments of a piecewise-chain.  The Gay-Berne potential accounts for the geometrical anisotropy of the interacting elements and uses a center-to-center distance, but this model can fail for elements with especially large aspect ratios which includes biopolymers such as actin \cite{gay1981modification, berardi1998gay}.  One alternative approach has been to use the closest distance between the two segments, but this can introduce discontinuities impairing simulation stability  \cite{sirk2012enhanced,kim2009computational}.  Intuitively, the interaction between two linear segments should arise as the integrated effect of the point-wise interactions between all pairs of points on the segments.  One can imagine subdividing the linear segments to numerically approximate this type of interaction, a method implemented in the ASPHERE package of LAMMPS \cite{plimpton1995fast}. However, by introducing more sampling points this approach negates the gain in efficiency from coarse-graining of the polymer into linear segments in the first place, as discussed below.  

The novel solution to these issues used in the MEDYAN model is to derive an analytical expression for the integrated effect of power-law repulsion between each differential element of the two interacting linear segments \cite{popov2016medyan}.  
A suitably steep power law function of the separation $r$ can be used as a smooth mimic of the hard-wall interaction.  Specifically, in MEDYAN the function $1/r^4$ serves as the interaction kernel of the double integral over the lengths of the two segments (see Equation \ref{eq1} below).  However, the result of the integration is an opaque and  complicated expression, and it contains degeneracies when the two linear segments are coplanar, leading to undefined behavior.  To address these shortcomings, in this paper we first clarify the calculation of the excluded volume repulsion potential used in MEDYAN.  Then we illustrate how the problem can be solved in the coplanar case and in other lower-dimensional geometries, and derive further expressions for the alternative steeper interaction kernel $1/r^{6}$.  We then characterize the dependence of these interactions on the configurations of the two segments, discuss how to parameterize the potential, and compare it to the widely-used Gay-Berne form.  We also introduce a new `segmental Lennard-Jones' interaction which has both attractive and repulsive components.  Finally, we implement a numerical approximation method and discuss the gain in computational efficiency from using the analytical expressions.

\section{Energies and Forces of the Integrated Interaction}

Here we derive analytical expressions for the excluded volume repulsion energy between two polymer segments, using the $1/r^4$ interaction kernel which corresponds to the implementation in MEDYAN.  We first give the derivation for cylindrical segments in 3D space.  Because the 3D expressions for the repulsion energy are not defined when the cylinders are coplanar, we next describe the steps for re-deriving these expressions in 2D scenarios.  Finally, we extend the derivations to apply to an interaction kernel of $1/r^6$, representing an even steeper hard-wall mimic.  Throughout this section certain complicated integrals must be solved, for which we use the computer algebra system (CAS) Mathematica \cite{mathematica, rich2018rule}.  We provide Mathematica notebook (.nb) files in the Supplementary Material which implement the calculations described below.   

\subsection{Segments in 3D}

\begin{figure}[H]
	\begin{center}
	\includegraphics[width=7cm]{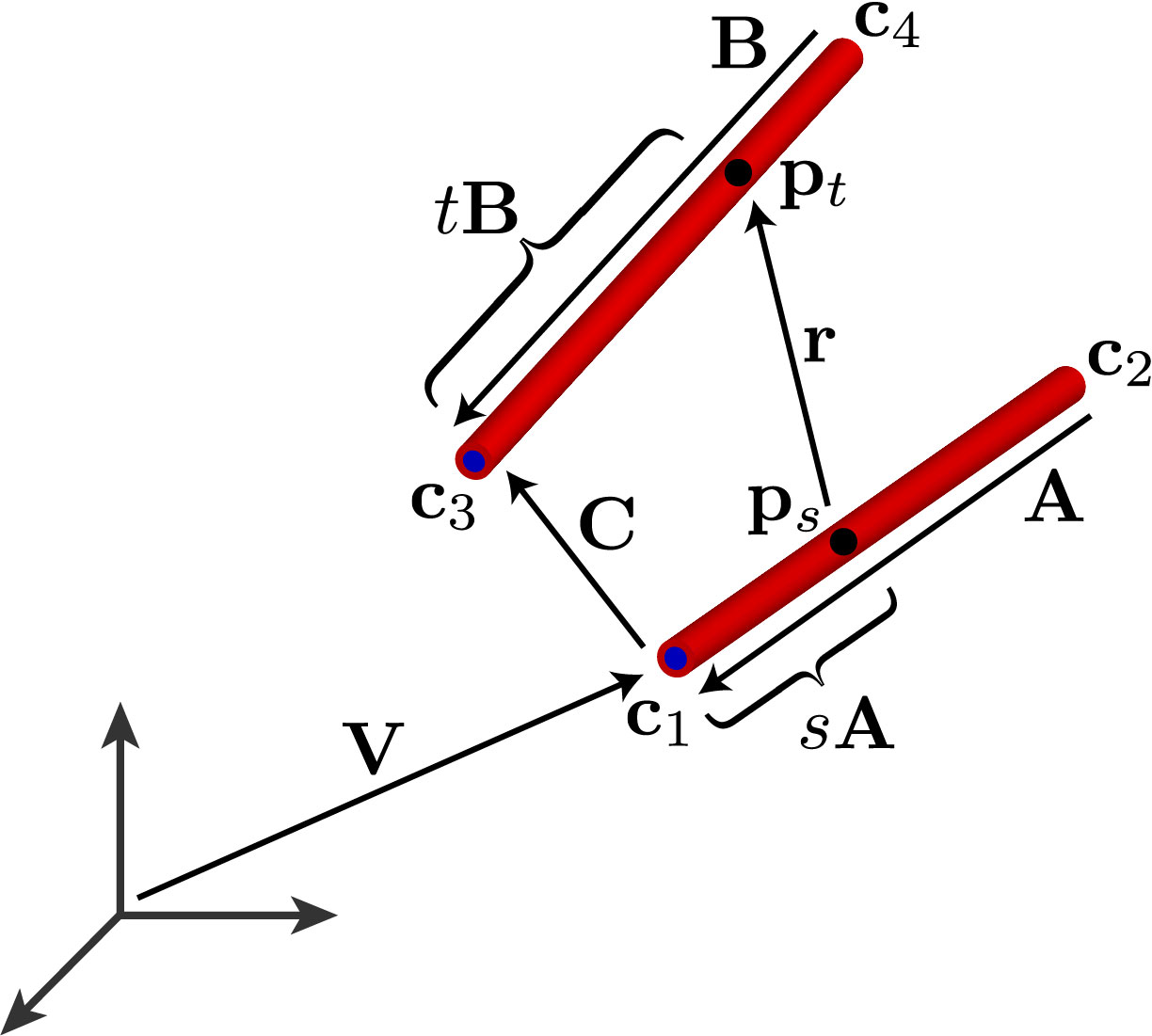}
	\caption{Two cylinders, $A$ and $B$, are in red, with the minus ends marked by blue dots.  $\mathbf{r}$ is defined by the points $\mathbf{p}_s$ and $\mathbf{p}_t$ on the cylinders.}
	\label{fig1}
\end{center}
\end{figure}

Given the positions and orientations of two thin cylinders, we define the excluded volume repulsion energy $U$ as proportional to a double integral of the function $1/r(s,t)^{4}$, where the integrals run over the length of each cylinder:
\begin{equation}
U = K_\text{vol} \int_0^1 \int_0^1 ds dt \frac{1}{r(s,t)^4}.
\label{eq1}
\end{equation}
Here $s$, $t$ $\in [0,1]$ parameterize the distance along the two cylinders $A$ and $B$ respectively, $r(s,t)$ denotes the magnitude of the vector $\mathbf{r}(s,t)$ which separates points on the two cylinder axes $\mathbf{p}_s$ and $\mathbf{p}_t$, and $K_\text{vol}$ is the constant of proportionality, having units of energy times length squared.  The geometry of the problem is illustrated in Figure \ref{fig1}.

The position and orientation of the two cylinders $A$ and $B$ can be specified by four vectors pointing to the positions of the four ends: $\mathbf{c}_1$ and $\mathbf{c}_2$ point to the minus and plus ends, respectively, of cylinder $A$ and $\mathbf{c}_3$ and $\mathbf{c}_4$ do likewise for cylinder $B$.  Alternatively, we can describe the two cylinders by the vectors
\begin{align}
\mathbf{V} &= \mathbf{c}_1 \\
\mathbf{A} &= \mathbf{c}_1 - \mathbf{c}_2 \\
\mathbf{B} &= \mathbf{c}_3 - \mathbf{c}_4 \\
\mathbf{C} &= \mathbf{c}_3 - \mathbf{c}_1.
\end{align}
We can represent a point on cylinder $A$ parameterized by $s$, $\mathbf{p}_s$, as 
\begin{equation}
\mathbf{p}_s = \mathbf{V} - s \mathbf{A},
\end{equation}
and similarly for a point on cylinder $B$ parameterized by $t$, $\mathbf{p}_t$, we have
\begin{equation}
\mathbf{p}_t = \mathbf{V} + \mathbf{C} - t \mathbf{B}.
\end{equation}
To solve the integral in Equation \ref{eq1}, first we need to write $r(s,t)$.  We have
\begin{equation}
\mathbf{r} = \mathbf{p}_t - \mathbf{p}_s = \mathbf{C} -t \mathbf{B} + s \mathbf{A}
\end{equation} 
and 
\begin{align}
r  =& \big(\mathbf{C} \cdot \mathbf{C} +2 s \mathbf{A} \cdot \mathbf{C}  +
s^2 \mathbf{A} \cdot \mathbf{A} \nonumber \\
&- 2 t \mathbf{B} \cdot \mathbf{C} - 2 s t \mathbf{A} \cdot \mathbf{B} + t^2 \mathbf{B} \cdot \mathbf{B}\big)^{1/2}. \label{req}
\end{align}
To simplify notation, we introduce the following variables:
\begin{align}
a &= \mathbf{A} \cdot \mathbf{A} \nonumber\\
b &= \mathbf{B} \cdot \mathbf{B} \nonumber\\
c &= \mathbf{C} \cdot \mathbf{C} \nonumber\\
d &= \mathbf{A} \cdot \mathbf{B} \nonumber\\
e &= \mathbf{A} \cdot \mathbf{C} \nonumber\\
f &= \mathbf{B} \cdot \mathbf{C} \nonumber.
\end{align}
Defining these intermediary variables is also computationally efficient by avoiding repeatedly calculating the same expressions.  With this, the goal is do the following integral:
\begin{align}
U =& K_\text{vol} \int_0^1 \int_0^1ds dt  \nonumber \\
&\frac{1}{(c + 2 e s + a s^2 - 2 f t - 2 d s t + b t^2)^2}.
\end{align}
This integral can be done with the help of a CAS resulting in a lengthy expression provided in the Appendix (Equation \ref{eqUlong}).  It can be  cleaned up somewhat by introducing the following variables:
\begin{align}
AA &= \sqrt{a c - e^2} \nonumber\\
BB &= \sqrt{b c - f^2} \nonumber\\
CC &= d e - a f \nonumber\\
DD &= b e - d f \nonumber\\
EE &= \sqrt{a(b+c-2f) - (d-e)^2} \nonumber\\
FF &= \sqrt{b(a+c+2e) - (d+f)^2} \nonumber\\
GG &= d^2 - ab - CC \nonumber\\
HH &= CC + GG - DD \nonumber\\
JJ &= c(GG+CC) + e DD - f CC \nonumber\\
ATG_1 &= \tan^{-1} \left( \frac{a+e}{AA} \right)  - \tan^{-1}\left(\frac{e}{AA}\right) \nonumber\\
ATG_2 &= \tan^{-1} \left( \frac{a+e-d}{EE} \right)  - \tan^{-1}\left(\frac{e-d}{EE}\right) \nonumber\\
ATG_3 &= \tan^{-1} \left( \frac{f}{BB} \right)  - \tan^{-1}\left(\frac{f-b}{BB}\right) \nonumber\\
ATG_4 &= \tan^{-1} \left( \frac{d+f}{FF} \right)  - \tan^{-1}\left(\frac{d+f-b}{FF}\right) \nonumber.\\
\nonumber
\end{align}  
These variable names are chosen to match those in the MEDYAN codebase.  We point out that the multiple letters comprising these variables do not indicate multiplication of two or more variables.  With this, the result is
\begin{align}
U =& \frac{K_\text{vol}}{2 JJ }\bigg(ATG_1 \frac{CC}{AA} + ATG_2 \frac{GG}{EE} \nonumber \\
&  +ATG_3 \frac{DD}{BB} +ATG_4 \frac{HH}{FF} \bigg).
\label{eqUshort}
\end{align}

Implementing excluded volume repulsion in simulation usually also requires expressions for the derivatives of the energy with respect to the cylinder endpoints $\mathbf{c}_i$, which are used to determine the forces for time integrator-based approaches (e.g. CytoSim) or equivalently the gradients for minimization-based approaches (e.g. MEDYAN).  Derivatives such as $\frac{\partial U}{\partial \mathbf{c}_1}$ can be found using the chain rule:
\begin{align}
    \frac{\partial U}{\partial \mathbf{c}_1} = \frac{\partial U}{\partial a}\frac{\partial a}{\partial \mathbf{c}_1} + \frac{\partial U}{\partial b}\frac{\partial b}{\partial \mathbf{c}_1} + \dots,
\end{align}
where the derivatives $\frac{\partial U}{\partial a}$, $\frac{\partial U}{\partial b}, \ \dots$, can be obtained from Equation \ref{eqUlong} in the Appendix, and the derivatives $\frac{\partial a}{\partial \mathbf{c}_i}$ can be found using the definitions given above.  For example,
\begin{align}
    \frac{\partial a}{\partial \mathbf{c}_1} = &  \frac{\partial}{\partial \mathbf{c}_1} ( \mathbf{c}_1 \cdot \mathbf{c}_1 - 2 \mathbf{c}_2 \cdot \mathbf{c}_1 + \mathbf{c_2} \cdot \mathbf{c}_2 ) \nonumber \\
    =& \mathbf{c}_1 - 2 \mathbf{c}_2.
\end{align}

\subsection{Segments in 2D}
When the cylinders $A$ and $B$ are in the same plane then the vectors $\mathbf{A}$, $\mathbf{B}$, and $\mathbf{C}$ are all coplanar and the problem becomes effectively 2D.  Some implementations may also simply assume a 2D space.  In this scenario, the scalar triple product $(\mathbf{A}\times \mathbf{B}) \cdot \mathbf{C}$ vanishes.  It can be shown by straightforward algebraic rearrangement that the quantity $JJ$ appearing in the denominator of the right hand side of Equation \ref{eqUshort} is given by
\begin{equation}
    JJ = -\left( (\mathbf{A}\times \mathbf{B}) \cdot \mathbf{C} \right)^2.
\end{equation}
As a result, the above expressions for the energy $U$ and derivatives $\frac{\partial U}{\partial \mathbf{c}_i}$ are not defined, and a special case must be considered.  

It is instructive to count the number of free variables in the 3D and 2D case.  In both settings, $U$ is invariant with respect to a rigid rotation or translation of the system.  In 3D, we originally have 12 variables (the 12 components of $\mathbf{c}_1$, $\mathbf{c}_2$, $\mathbf{c}_3$, and $\mathbf{c}_4$), but translation invariance implies that 3 degrees of freedom are extraneous and rotation invaraince implies that 3 additional degrees of freedom are extraneous.  This leaves 6 independent degrees of freedom, which appear in the expression for $U$ as $a$, $b$, $c$, $d$, $e$, $f$.  In 2D, we originally have $8$ degrees of freedom, but translation invariance implies that 2 degrees of freedom are extraneous and rotation invariance implies than additional 1 degree of freedom is extraneous, leaving 5 degrees of freedom.  Indeed, the condition in 2D that $JJ = 0$ implies an additional constraint among the 6 variables.  It can be shown that, in the 2D case but not in the 3D case,
\begin{equation}
    f = \frac{d e + \sqrt{(ab-d^2)(ac - e^2)}}{a},
\end{equation}
and hence only 5 variables are free in 2D.  

One could through substitution write the integrand $1/r^4$ in terms of five free variables in 2D, but this becomes an algebraically complicated expression that precludes exact integration.  Instead, we first rotate the configuration so the shared plane coincides with the $xy$ plane.  Next, we write the integrand using the 6 (redundant) vector components $A_x, \ A_y, \ B_x, \ B_y, \ C_x$, and $ C_y$ to find the interaction energy in 2D.  We have \begin{align}
   U =&  K_\text{vol} \int_0^1 \int_0^1 ds dt \nonumber \\
   &\frac{1}{\left((C_x + A_x s - B_x t)^2 + (C_y + A_y s - B_y t)^2 \right)^2}.
\end{align} 
This integral has a complicated result which is provided in the Appendix (Equation \ref{coplanar}).  The denominator of the result is proportional to $A_y B_x-A_x B_y$.  If $\mathbf{A}$ and $\mathbf{B}$ are parallel (or anti-parallel) in addition to coplanar, then one can show that $A_x B_y = A_y B_x$, and hence the expression for $U$ in the coplanar case is not defined.  

When $\mathbf{A}$ and $\mathbf{B}$ are (anti-)parallel, then $\mathbf{B} = \xi \mathbf{A}$ for some $\xi \in \rm I\!R$, $\xi \neq 0$.  Expressing the integrand using this new variable, we have
\begin{align}
   U =&  K_\text{vol} \int_0^1 \int_0^1 ds dt \nonumber \\
   &\frac{1}{\left((C_x + A_x (s - \xi t))^2 + (C_y + A_y (s - \xi t))^2 \right)^2}.
\end{align} 
The result of this integral is also provided in the Appendix (Equation \ref{parallel}).  The denominator of that result is proportional to $(A_y C_x - A_x C_y)^3$.  If, in addition to being parallel, $\mathbf{A}$ and $\mathbf{B}$ are colinear, then $\mathbf{C}$ is parallel to $\mathbf{A}$ and $A_x C_y = A_y C_x$, and this result is not defined.  

When $\mathbf{A}$ and $\mathbf{B}$ are colinear, one may write $\mathbf{C} = \zeta \mathbf{A}$ and express the integrand as 
\begin{align}
   U =&  K_\text{vol} \int_0^1 \int_0^1 ds dt \nonumber \\
   &\frac{1}{\left((A_x (\zeta + s - \xi t))^2 + (A_y (\zeta + s - \xi t))^2 \right)^2}.
\end{align} 
The result of this integral, also provided in the Appendix (Equation \ref{colinear}), is simpler than in the previous cases, depending on just 4 variables $A_x, \ A_y, \ \xi,$ and $\zeta$.  Several ratios appear in the result with denominators proportional to $(A_x^2 + A_y^2)^2, \ \xi, \ \zeta, \ (\zeta - \xi)^2, \ (\zeta - \xi + 1)^2$, and $(\zeta + 1)^2$.  Requiring that none of these are zero implies that $\mathbf{A}$, $\mathbf{B}$, and $\mathbf{C}$ are all non-zero, and that $\mathbf{A}$ and $\mathbf{B}$ are nowhere coincident in which case the interaction would diverge.  

The treatment given above for handling the special case scenarios of cylinder configurations that lead to degeneracies in the energy expressions is not exhaustive, and certain degeneracies remain (such as one coming from $A_xC_x + A_y C_y = 0$ in the parallel case, Equation \ref{parallel}).  These degeneracies, unlike those coming from geometrical properties of the configurations, remain as a result of expressing the integrand using redundant variables (which was necessary to do the integration analytically).  It would be straightforward to consider each degeneracy in turn and, by following steps similar to those outlined above, derive backup expressions for each scenario.  However, we recommend instead implementing a numerical approximation method to fall back on when these degeneracies are encountered in simulation.  Such a numerical method is described below.  We emphasize that in a given dimensionality, the manifold of cylinder configurations leading to degeneracies is of lower dimension than the ambient space and hence such configurations will be exceedingly rare under typical physical dynamics.

\subsection{$1/r^6$ interaction kernel}
In Equation \ref{eq1}, the repulsion energy between two cylinders was taken as a double integral over both cylinder lengths of the interaction kernel $1/r^4$.  The $1/r^4$ interaction is fairly steep, mimicking a hard-wall boundary with an effective cylinder radius set by the choice of prefactor $K_\text{vol}$.  However one may prefer an even steeper potential than $1/r^4$, such as $1/r^6$, so that the range of separation over which the interaction starts to be felt is narrower.  The new interaction energy is expressed as
\begin{equation}
U = K_\text{vol} \int_0^1 \int_0^1 ds dt \frac{1}{r(s,t)^6}.
\label{eq6}
\end{equation}
Such a potential mimics even more closely a true hard-wall interaction (see Figure \ref{param} below), and with both interactions in hand it becomes possible by combining them to create bimodal energy profiles, similar to a Lennard-Jones potential.  It is straightforward to carry through identical steps for the $1/r^6$ kernel as outlined above for the $1/r^4$ kernel, with the same issues of degeneracies arising from special-case cylinder configurations.  For brevity, and since no new concepts are involved, we skip the discussion here of how those steps are carried out and also omit the resulting expressions from the Appendix.  The expressions can be found in the supplementary Mathematica notebook files.

\section{Examples and Parameterization}
\subsection{Comparing endpoint-based and integrated kernel interactions}
Here we analyze the `integrated kernel' energy functions (Equations \ref{eq1} and \ref{eq6}) and discuss notable features arising from a set of test cases.  For comparison, we also introduce two other `endpoint-based' interaction functions which, rather than integrating the kernels $1/r^4$ or $1/r^6$ over the lengths of the cylinders, simply include repulsion between felt by the endpoints of the two cylinders:
\begin{align}
    U = K_\text{vol} \bigg( & \frac{1}{m(\mathbf{c}_3,\mathbf{c}_1,\mathbf{c}_2)^4 }  +\frac{1}{m(\mathbf{c}_4,\mathbf{c}_1,\mathbf{c}_2)^4  }  + \nonumber \\
    & \frac{1}{m(\mathbf{c}_1,\mathbf{c}_3,\mathbf{c}_4)^4  } + \frac{1}{m(\mathbf{c}_2,\mathbf{c}_3,\mathbf{c}_4)^4  } \bigg) \label{eqdisc4}
\end{align}
and
\begin{align}
    U = K_\text{vol} \bigg( & \frac{1}{m(\mathbf{c}_3,\mathbf{c}_1,\mathbf{c}_2)^6 }  +\frac{1}{m(\mathbf{c}_4,\mathbf{c}_1,\mathbf{c}_2)^6  }  + \nonumber \\
    & \frac{1}{m(\mathbf{c}_1,\mathbf{c}_3,\mathbf{c}_4)^6  } + \frac{1}{m(\mathbf{c}_2,\mathbf{c}_3,\mathbf{c}_4)^6  } \bigg) \label{eqdisc6}
\end{align}
where $m(\mathbf{c}_x,\mathbf{c}_a,\mathbf{c}_b)$ represents the minimal distance from the point $\mathbf{c}_x$ to the line segment connecting $\mathbf{c}_a$ and $\mathbf{c}_b$.  Expressions of this type are sometimes used to model the steric repulsion of polymers, but we show below that they have the deficiency of a relatively flat energy profile for cylinder separations much less than the cylinder length, which can allow cylinders to overlap each other under typical dynamics \cite{rickman2019effects,letort2015geometrical}.  

\begin{figure}[H]
	\begin{center}
	\includegraphics[width=13 cm]{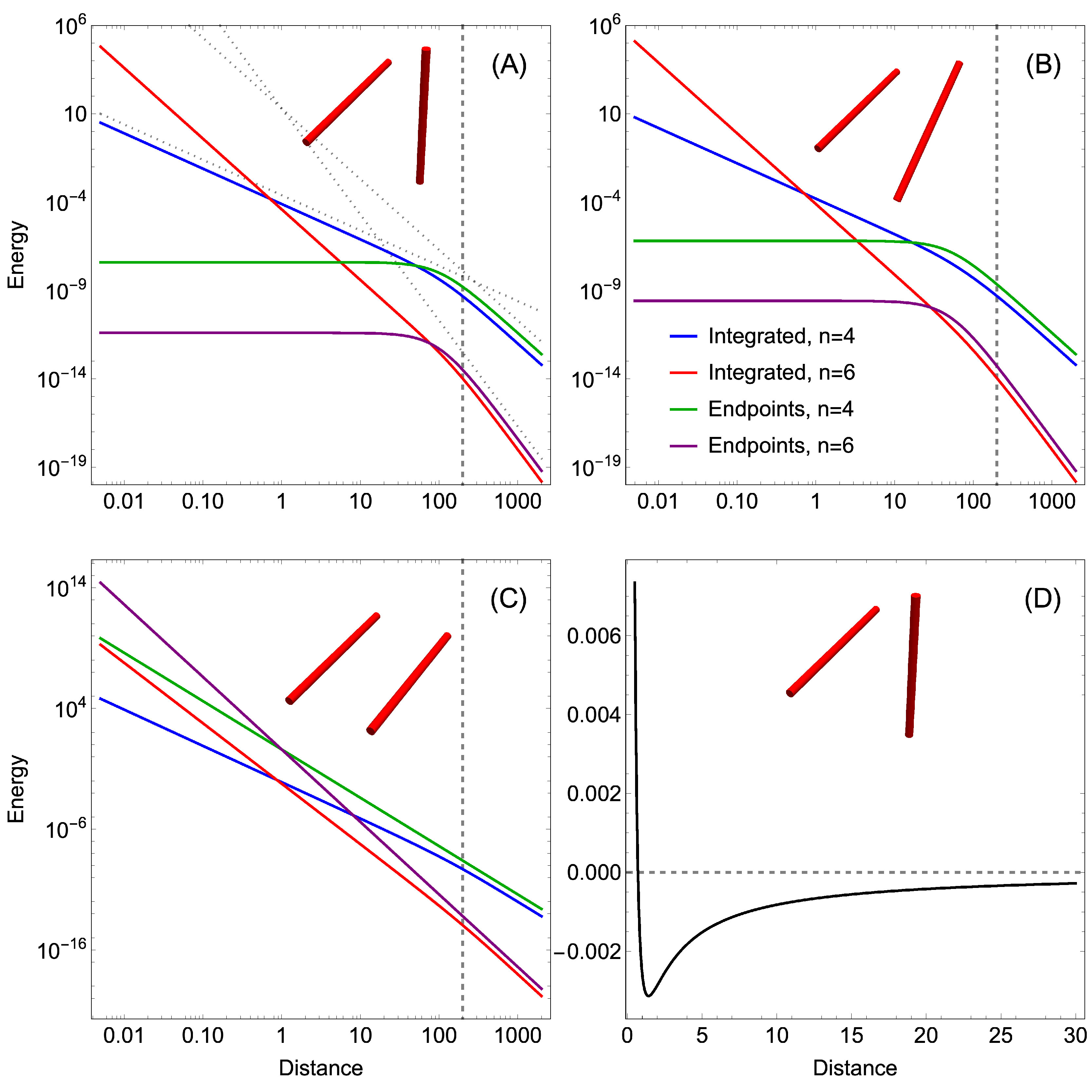}
	\caption{Plots of interaction energy are shown as the inter-cylinder distance $l$ is varied for different cylinder orientations and choices of interaction energy.  Units are arbitrary, as length and energy scales are set by the $K_\text{vol}$ prefactor which is taken to be unity (amounting to setting the vertical position of each curve on the log-log plots).  (A) The blue line represents Equation \ref{eq1}, the red line represents Equation \ref{eq6}, the green line represents Equation \ref{eqdisc4}, and the purple line represents Equation \ref{eqdisc6}.  The dotted vertical line indicates the length of the two cylinders.  The cylinders are oriented perpendicularly to each other as visualized in the inset.  The light dotted lines proportional to $1/l^2$, $1/l^4$, and $1/l^6$ show the scaling behaviors in different regimes.  (B) The same plot is shown as in panel (A), except for a relative orientation of $30^\circ$ between the cylinders.  (C) The same plot is shown as in panel (A), except for a relative orientation of $0^\circ$ between the cylinders.  (D) A plot is shown of a segmental Lennard-Jones interaction energy profile from combining the $1/r^4$ and $1/r^6$ integrated kernel energies (Equation \ref{eqhybr}) .  The horizontal dotted line separates the attractive and repulsive regions.}
	\label{mpf}
\end{center}
\end{figure}

We consider two cylinders each of length $200$, where the units are fixed by setting $K_\text{vol} = 1$ throughout.  The cylinders are aligned (i.e. $\mathbf{A} \cdot \mathbf{C} = 0$), and the vector $\mathbf{l}$ joining the each cylinder's midpoint is kept perpendicular to each cylinder as we rotate one cylinder about this vector, producing different relative configurations.  This set-up is visualized in the insets of Figure \ref{mpf}.A-C.  For each configuration we vary the distance $l$ (the magnitude of $\mathbf{l}$), and study the effect on the various interaction energies Equations \ref{eq1}, \ref{eq6}, \ref{eqdisc4}, and \ref{eqdisc6}.  The results are displayed in Figure \ref{mpf}.

Two key features are evident from this example.  First,  there are different asymptotic behaviors in the small and large distance regimes, with a crossover around distances on the order of $L$.  For $l \ll L$, the integrated expressions behave like a power law with an exponent equal to that of the kernel function plus 2, whereas for $l \gg L$, the integrated expressions behave like a power law with an exponent equal to that of the kernel function:
\begin{equation}
    U = \int_0^1 \int_0^1 ds dt \frac{1}{r(s,t)^n} \sim \begin{cases} 
      l^{2-n} & l \ll L \\
      l^{-n} & l \gg L
   \end{cases}
\end{equation}
where $n=4$ and $6$ in the examples shown.  This behavior is expected, since in the far field all points in the cylinders repel each other with similar magnitudes, whereas in the near field the repulsion is dominated only by nearby points, changing the scaling by a factor of $l^2$.  Second, we see that for the non-coplanar cylinder configurations (Figures \ref{mpf}.A and \ref{mpf}.B), the endpoint-based interaction energies have flat energy profiles for distances much less than the cylinder length.  For the endpoint-based functions, 
\begin{equation}
    U \sim \begin{cases} 
      1 & l \ll L \\
      l^{-n} & l \gg L
   \end{cases}.
\end{equation}
One can understand this as resulting from the fact that, when $l \ll L$, the distance between the endpoints of non-coplanar cylinders change much less than the distance between points the middle of the cylinder as $l$ is decreased, and these points in the middle contribute do not contribute to the energy penalty in the endpoint-based case.  This qualitative difference between the integrated and endpoint-based interactions is much less pronounced in the coplanar case, when the distance between the endpoints change at the same rate as all points when $l$ is decreased, as shown in Figure \ref{mpf}.C.  

We next introduce a `segmental Lennard-Jones' interaction potential
\begin{align}
U =& K_\text{vol,6} \left( \int_0^1 \int_0^1 ds dt \frac{1}{r(s,t)^6} \right)^{{n_6}}  \nonumber \\
& -K_\text{vol,4} \left( \int_0^1 \int_0^1 ds dt \frac{1}{r(s,t)^4} \right)^{{n_4}}.
\label{eqhybr}
\end{align}
This expression has qualitative similarity to the familiar 6-12 Lennard-Jones potential between two particles, and can be tuned by choosing the four parameters $K_\text{vol,6}$, $K_\text{vol,4}$, $n_4$, and $n_6$ to mimic interactions that have both attractive and repulsive parts.  For instance, computational modeling of depletion forces, which tend to aggregate polymers together, may make use of an effective attractive component in the polymer-polymer interaction \cite{letort2015geometrical}.  An example of a segmental Lennard-Jones potential is illustrated in Figure \ref{mpf}.D for the choices $K_\text{vol,6}=K_\text{vol,4}=1$ and $n_4 = n_6 = 1/2$ and for a perpendicular configuration of the cylinders.

\end{multicols}
\begin{multicols}{2}

\subsection{Determination of $K_\text{vol}$}
\begin{figure}[H]
	\begin{center}
	\includegraphics[width=7 cm]{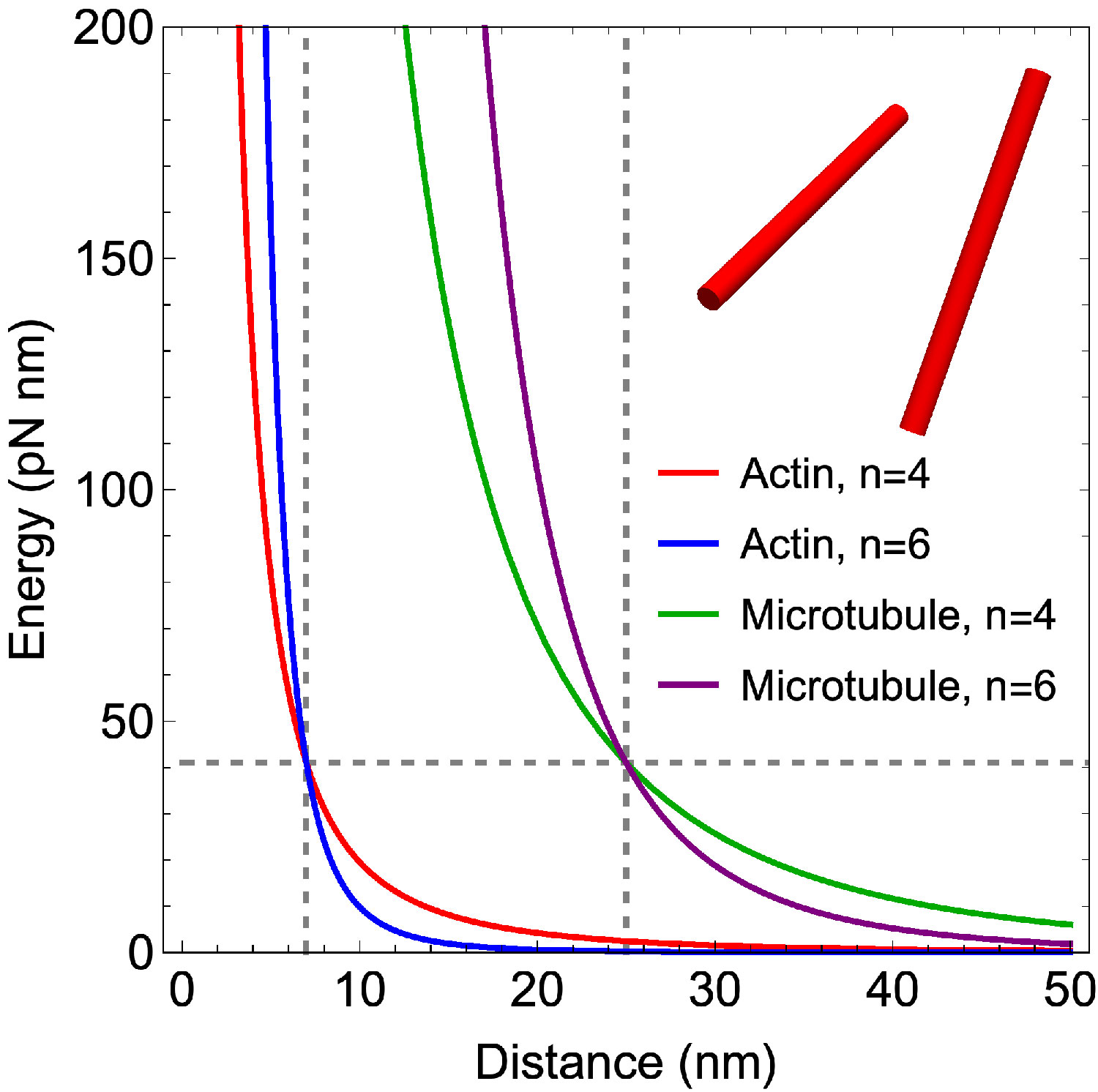}
	\caption{Plots of the energy as a function of the separation distance are shown for different parameterizations and interaction kernels.  The red and blue curves are parameterized for actin using the $1/r^4$ and $1/r^6$ kernel respectively, while the green and purple curves are likewise paramaterized for microtubules.  The dashed vertical lines are drawn at $d^*$ for actin and microtubules, and the intersection points indicate the enforcement of the condition $U_t(d^*) = U_m$ where $U_m = 41$ pN nm (horizontal dahsed line).}
	\label{param}
\end{center}
\end{figure}

Actual biopolymers can differ significantly in their diameters, requiring that $K_\text{vol}$ be tuned for particular biopolymers.  This choice can be made so that at the effective diameter $d^*$ the typical interaction energy $U_t$ (i.e. the energy for some typical configuration of segments) is equal to some energetic penalty for steric overlap $U_m$ of the system:
\begin{equation}
    U_t(d^*) = U_m.
    \label{eqparam}
\end{equation}
Here we indicate how to use Equation \ref{eqparam} to determine $K_\text{vol}$ for actin and microtubules which have been modeled as chains of $100$ nm long linear segments.  The radius of an actin filament is approximately $3.5 \ nm$, and for a microtubule it is $12.5$ nm.  As a typical configuration we take the two interacting segments to be aligned and rotated by $45^\circ$ with respect to each other.  In Figure \ref{param} we show the result of using $U_t(d^*) = U_m$ to determine $K_\text{vol}$, taking $U_m = 41$ pN nm to be 10 times the thermal energy (as $k_B T = 4.1$ pN nm at room temperature).  For actin segments, this procedure gives $K_\text{vol} = 4.6 \times 10^6$ pN nm$^3$ for the $1/r^4$ kernel and $K_\text{vol} = 4.4 \times 10^8$ pN nm$^5$ for the $1/r^6$ kernel.  For microtubule segments, this procedure gives $K_\text{vol} = 7.7 \times 10^7$ pN nm$^3$ for the $1/r^4$ kernel and $K_\text{vol} = 7.7 \times 10^{10}$ pN nm$^5$ for the $1/r^6$ kernel.  We note that this parameterization process also depends on the chosen length of the cylindrical segments.

\subsection{Sensitivity of $K_\text{vol}$ to typical configurations}
\begin{figure}[H]
	\begin{center}
	\includegraphics[width=8 cm]{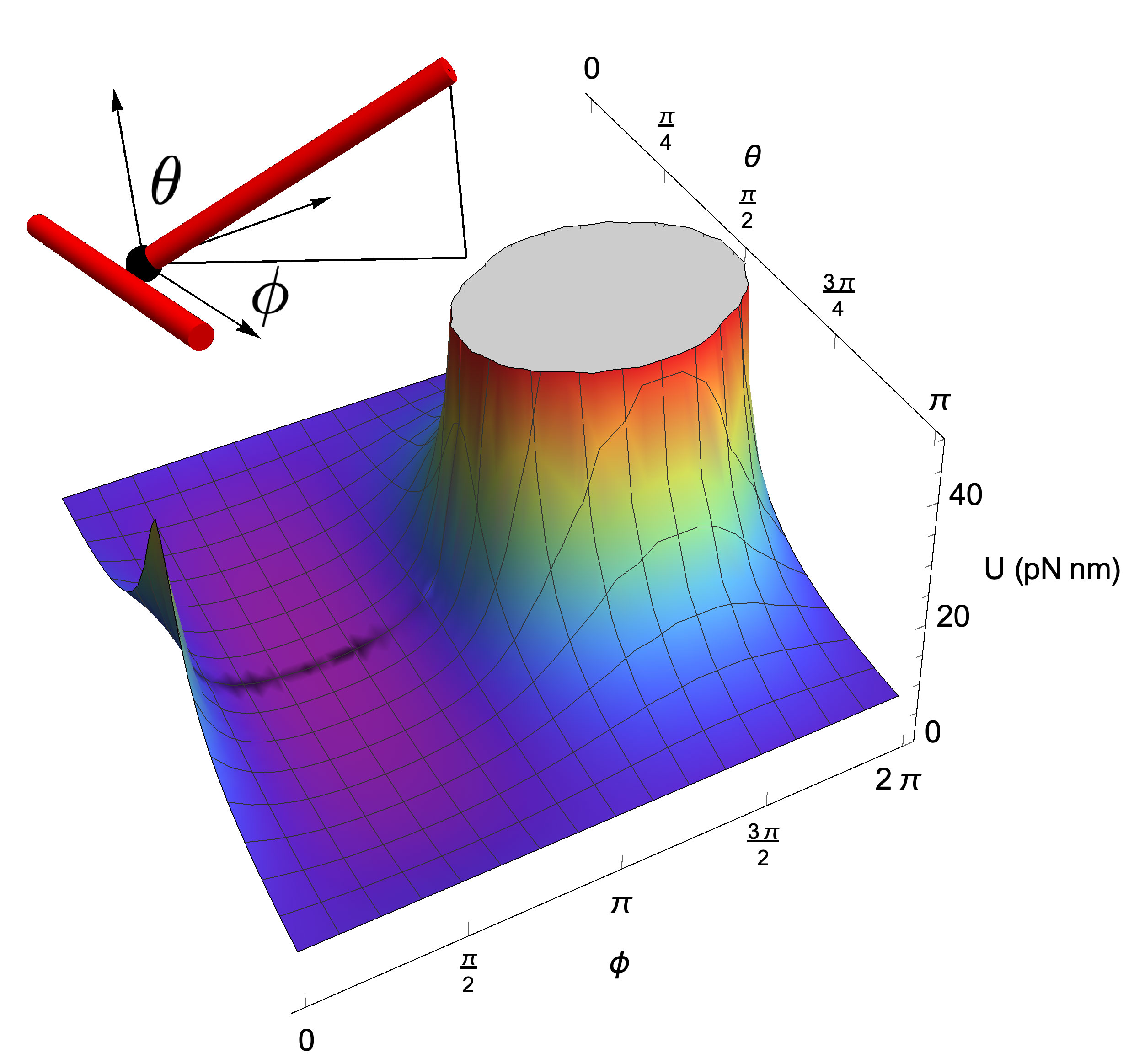}
	\caption{A surface plot of $U$ as $\theta$ and $\phi$ are varied is shown for the conditions described in the main text.  The colors indicate the energy, ranging from $0$ pN nm (purple) to 5$0$ pN nm (red), the gray region corresponds to configurations in which the segments nearly overlap, and the inset illustrates the definition of the spherical coordinates $\theta$ and $\phi$.  }
	\label{angles}
\end{center}
\end{figure}

The parameterization method described above has one seemingly major ambiguity, which is how to determine the typical configuration of the segments at which to evaluate $U_t(d^*)$.  Fixing the position of one segment and both segments' lengths, 5 variables remain to specify the other segment: the offset vector $\mathbf{C}$ and the spherical coordinates $\theta$ (inclination) and $\phi$ (azimuthal) of the unit vector $\hat{\mathbf{B}}$.  The dependence on the separation of aligned cylinders has been discussed above.  We next explore the orientational coordinates $\theta$ and $\phi$, setting
\begin{align}
    \mathbf{c}_1 &= (-L/2, \ -l, \ 0) \nonumber \\
    \mathbf{c}_2 &= (L/2, \ -l, \ 0) \nonumber \\
    \mathbf{c}_3 &= (0, \ 0, \ 0) \nonumber \\
    \mathbf{c}_4 &= (L \sin \theta \cos \phi, \ L \sin \theta \sin \phi, \ L \cos\theta) \nonumber. 
\end{align}
The geometry of this set up is illustrated in the inset of Figure \ref{angles}.

In Figure \ref{angles}, we show the energy $U$ for $L = 100$ nm, $l = 10$ nm, and $K_\text{vol} = 4.6 \times 10^6$ pN nm$^3$ as a function of $\theta$ and $\phi$.  Evidently, $U$ lies within a fairly small range over most of the domain of $\theta$ and $\phi$ (with the exception being where the cylinder nearly overlap).  It can similarly be shown that the degree of freedom corresponding to sliding one cylinder in a direction parallel to the other cylinder only affects the interaction energy to within an order of magnitude.  Therefore, we may conclude that the parameterization is not very sensitive to how the typical configuration is chosen, and that reliable order of magnitude estimates of $K_\text{vol}$ can be obtained for a given $d^*$ and $U_m$.

\subsection{Comparison to Gay-Berne potential}
\begin{figure}[H]
	\begin{center}
	\includegraphics[width=7 cm]{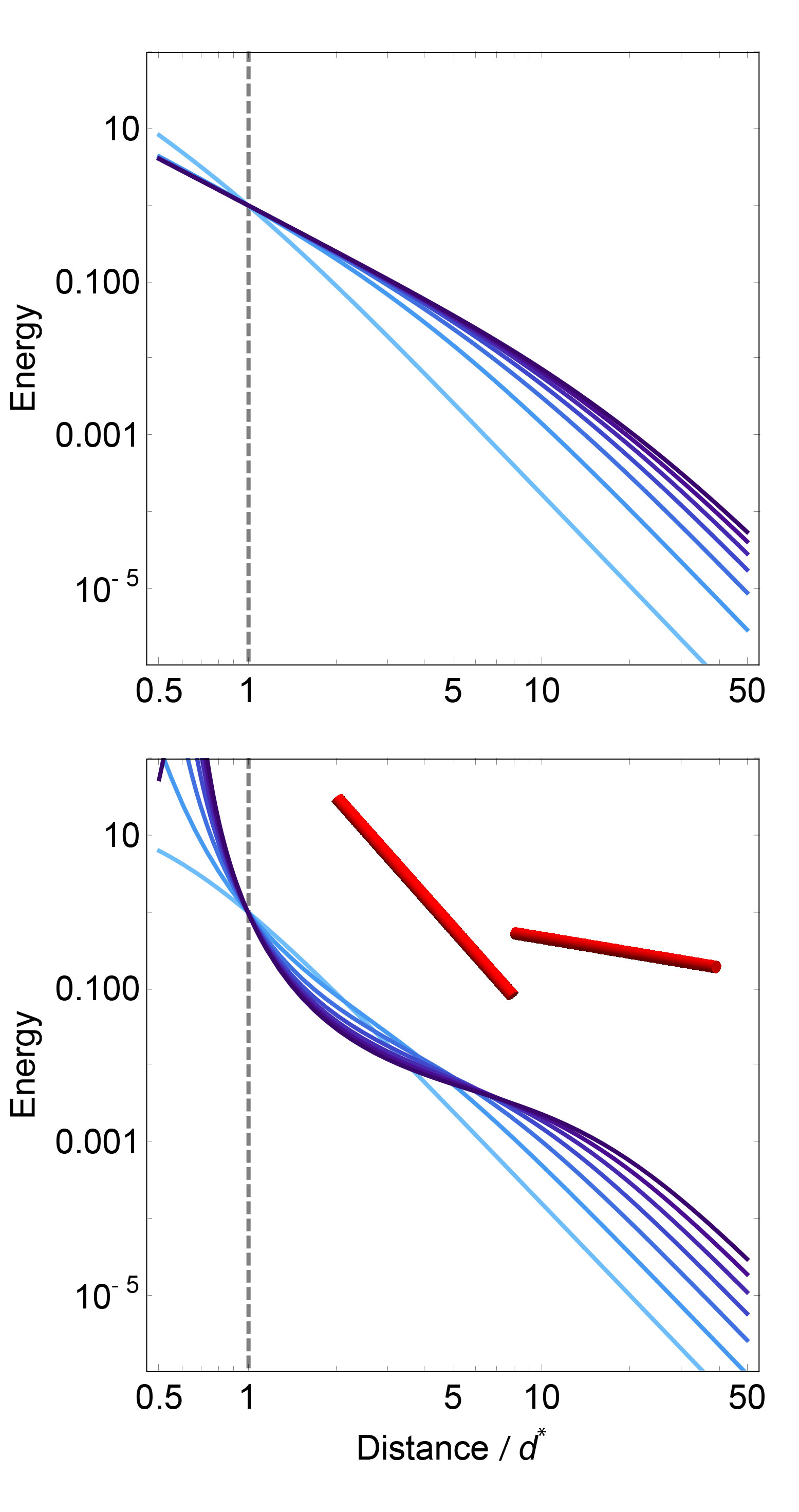}
	\caption{Interaction energy profiles for the integrated $1/r^4$ kernel potential (top panel, Equation \ref{eq1}) and Gay-Berne potential (bottom panel, Equation \ref{GBeq}), and the  are shown.  In both panels, the geometric aspect ratio $\kappa$ is varied in increments of $5$ from $1$ to $31$ (i.e. $L$ is increased with $d^*$ fixed), as the colors are varied from light blue to dark purple.  The inset shows the set-up of the two cylinders for $\kappa = 21$. }
	\label{GB}
\end{center}
\end{figure}

Next, we compare the new integrated kernel expression for the interaction energy between cylindrical objects, Equation \ref{eq1}, to the widely used Gay-Berne potential which describes the interaction between anisotropic ellipsoidal objects. The original Gay-Berne potential was designed to be similar to a Lennard-Jones potential, having both attractive and repulsive contributions, but for comparison here we modify the original potential to be only repulsive and with an exponent of $-4$; we give the formula for the Gay-Berne potential used here in the Appendix (Equation \ref{GBeq}).  We compare the interaction energy profiles as a function of distance for two offset, rotated cylindrical segments with variable aspect ratios.  Holding the diameters $d^* = 1$ fixed, we change the cylinder lengths $L = \kappa d^*$, where $\kappa$ is the geometric aspect ratio, and show that for large $\kappa$ the Gay-Berne potential deviates strongly from the desired power-law repulsion.  We use the following test case configuration, illustrated in the inset of Figure \ref{GB}: for each choice of $L$, the horizontal offset (along their lengths) of two parallel, initially aligned cylinders is chosen such that half of their lengths overlap, and one cylinder is then rotated $45^\circ$ around around the line joining its midpoint and the other cylinder's overlapping endpoint.  The length of this line is then varied to construct the interaction energy profile for this test configuration.  The energy scale is fixed by setting each energy to $1$ at a distance of $d^*$.  We display the results in Figure \ref{GB}.

We observe that for large values of $\kappa$, the Gay-Berne interaction profile deviates significantly from the expected power-law behavior, exhibiting weakened repulsion for $l \gtrsim d^*$ and enhanced repulsion for $l \lesssim d^*$ compared to the integrated kernel interaction.  On the other hand, for $\kappa = 1$ the Gay-Berne profile and the integrated kernel profile nearly coincide.  

The need for a new potential to describe polymer repulsion can be understood as arising from the fact that, when modeling consecutive cylinders in a polymer as ellipsoids, the potential energy is not uniform along the polymer's length.  One can imagine a chain of sausage links to represent this scenario.  In the integrated kernel interaction however, the energy is uniform and therefore does not depend on how the polymer is discretized into cylinders.  This is an important physical feature to preserve in computational modeling.

\begin{figure}[H]
	\begin{center}
	\includegraphics[width=7 cm]{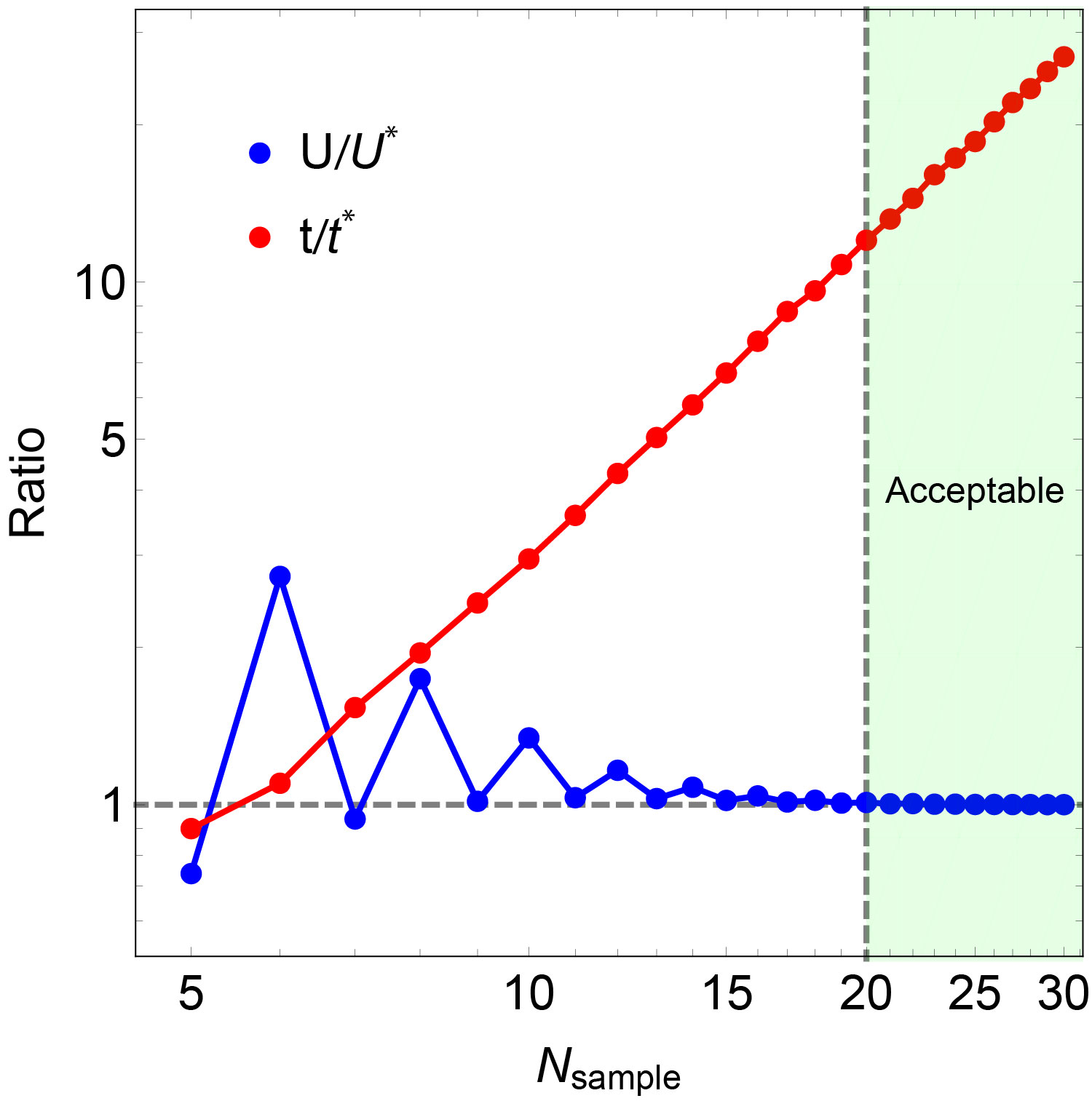}
	\caption{The ratio of the numerically obtained energy $U$ to the analytical $U^*$ and the ratio of the numerical evaluation time $t$ to that of the analytical result $t^*$ are shown as the number of sampling points $N_\text{sample}$ is varied.  The green shaded area indicates where the agreement between $U$ and $U^*$ is acceptable (i.e. $U/U^* \approx 1$).  The timing data is an average over $100$ repetitions.}
	\label{Timing}
\end{center}
\end{figure}

\subsection{Timing of Numerical Approximation}
Finally, we illustrate the gain in computational efficiency from having an analytical result (Equation \ref{eqUshort}) for the integral in Equation \ref{eq1} rather than a numerical approximation, as is sometimes used in LAMMPS \cite{plimpton1995fast}.  We implemented both the analytical result and a numerical scheme sampling the double integral at $N_\text{sample}$ points along each cylinder in compiled C code.  The numerical scheme approximates $U$ as 
\begin{equation}
U \approx \frac{K_\text{vol}}{N^2_\text{sample}}\sum_{i,j = 1}^{N_\text{sample}} \frac{1}{r\left(\frac{i}{N_\text{sample}},\frac{j}{N_\text{sample}}\right)^4},
\label{eqnum}
\end{equation}
where $r(s,t)$ is given in Equation \ref{req}.  For a single test case of aligned cylinders rotated by $45^\circ$ relative to each other, we compared the energy and evaluation time for the numerical scheme to the analytical counterpart as $N_\text{sample}$ was varied from $5$ to $30$.  The result is displayed in Figure \ref{Timing}.  Once the number of sampling points is large enough that the numerical approximation is acceptable ($N_\text{sample} \sim 20$), the numerical evaluation time is at least $10$ times longer than the evaluation time of Equation \ref{eqUshort}.  We note that to obtain a numerical approximation to the forces, the derivative with respect to the cylinder points $\mathbf{c}_i$ can be brought inside the sum in Equation \ref{eqnum}.

\section{Conclusion}
Our goal has been to clarify the derivation of the novel excluded volume repulsion potential implemented in MEDYAN and to extend the derivation to other scenarios of interest.  This overall approach to modeling repulsion interactions based on integrating an interaction kernel may be extended to other geometrical elements of finite size, such as 2D faces or 3D volumes (see Ref. \citenum{Ni2021membrane} for an application to 2D faces).  Despite the complexity of the resulting expressions for the energy and forces, they have the significant benefit of being analytical and avoiding endpoint-based interactions, which have flat energy profiles that can allow the repelling objects to erroneously overlap each other.  On the other hand, these expressions have the issue of being undefined for certain lower-dimensional rare configurations which impedes their usability in simulation.  However, we have shown how this issue can be handled by re-deriving expressions using a reduced number of variables.  In addition, we described how other types of interactions can be designed, such as steeper repulsion and a segmental Lennard-Jones interaction, while still accounting for the finite dimensions of the interacting objects.  This potential could be useful to model certain aggregating polymer systems such as toroidal DNA \cite{bloomfield1996dna,leforestier2009structure}.  The mathematical elaborations presented here should enable other investigators to effectively use these new potentials in their computational studies of soft matter systems.

\section*{Acknowledgements}
We thank Wonyeong Jung and Tae Yoon Kim for helpful discussions.  This work was supported by the grants 1632976 and CHE-1800418 from the National Science Foundation.  

\clearpage
\end{multicols}

\section*{Appendix}
Here we provide the full expressions of the interaction energy using the $1/r^4$ interaction kernel in various types of cylinder configurations.  The meaning of the variables is provided in the main text.
\subsection*{Cylinders in 3D}
\begin{align}
U = &\frac{K_\text{vol}}{2 \left(a \left(f^2-b c\right)+e (b e-2 d f)+c d^2\right)} \Bigg( \frac{\left(a b+b e-d^2-d f\right) \tan ^{-1}\left(\frac{-d-f}{\sqrt{a b+b (c+2
			e)-(d+f)^2}}\right)}{\sqrt{a b+b (c+2 e)-(d+f)^2}} \nonumber \\
			& - \frac{\left(a b+b e-d^2-d f\right)
	\tan ^{-1}\left(\frac{b-d-f}{\sqrt{a b+b (c+2 e)-(d+f)^2}}\right)}{\sqrt{a b+b (c+2
		e)-(d+f)^2}} +\frac{(a (b-f)+d (e-d)) \tan ^{-1}\left(\frac{e-d}{\sqrt{a (b+c-2
			f)-(d-e)^2}}\right)}{\sqrt{a (b+c-2 f)-(d-e)^2}}  \nonumber \\
			&+ \frac{(a (f-b)+d (d-e)) \tan
	^{-1}\left(\frac{a-d+e}{\sqrt{a (b+c-2 f)-(d-e)^2}}\right)}{\sqrt{a (b+c-2
		f)-(d-e)^2}} +\frac{\tan ^{-1}\left(\frac{e}{\sqrt{a c-e^2}}\right) (a f-d e)}{\sqrt{a
		c-e^2}} \nonumber \\
		& + \frac{\tan ^{-1}\left(\frac{a+e}{\sqrt{a c-e^2}}\right) (d e-a f)}{\sqrt{a
		c-e^2}} +\frac{\tan ^{-1}\left(\frac{b-f}{\sqrt{b c-f^2}}\right) (b e-d f)}{\sqrt{b
		c-f^2}} +\frac{\tan ^{-1}\left(\frac{f}{\sqrt{b c-f^2}}\right) (b e-d f)}{\sqrt{b
		c-f^2}} \Bigg) \label{eqUlong} 
\end{align}

\subsection*{Coplanar cylinders}

\begin{align}
    U &= \frac{K_\text{vol}}{4 \left(A_y B_x-A_x B_y\right)} \Bigg(\frac{\left(A_y B_x-A_x B_y\right){}^2}{\left(A_y \left(C_x-B_x\right)+A_x \left(B_y-C_y\right)\right) \left(A_y C_x-A_x C_y\right) \left(B_y C_x-B_x C_y\right)} \nonumber \\
    & +\frac{\left(A_y
   B_x-A_x B_y\right){}^2}{\left(A_y \left(C_x-B_x\right)+A_x \left(B_y-C_y\right)\right) \left(A_y C_x-A_x C_y\right) \left(A_y B_x+C_y B_x-A_x B_y-B_y C_x\right)}\nonumber \\
   & +\frac{\tan
   ^{-1}\left(\frac{A_x C_x+A_y C_y}{A_y C_x-A_x C_y}\right) \left(A_x^2+A_y^2\right)}{\left(A_y C_x-A_x C_y\right){}^2}-\frac{\tan ^{-1}\left(\frac{A_x^2+C_x A_x+A_y
   \left(A_y+C_y\right)}{A_y C_x-A_x C_y}\right) \left(A_x^2+A_y^2\right)}{\left(A_y C_x-A_x C_y\right){}^2} \nonumber \\
   &-\frac{\tan ^{-1}\left(\frac{B_x C_x+B_y C_y}{B_y C_x-B_x C_y}\right)
   \left(B_x^2+B_y^2\right)}{\left(B_y C_x-B_x C_y\right){}^2}+\frac{\tan ^{-1}\left(\frac{-B_x^2+C_x B_x+B_y \left(C_y-B_y\right)}{B_y C_x-B_x C_y}\right)
   \left(B_x^2+B_y^2\right)}{\left(B_y C_x-B_x C_y\right){}^2} \nonumber \\
   &+\frac{\tan ^{-1}\left(\frac{A_x B_x+C_x B_x+B_y \left(A_y+C_y\right)}{-A_y B_x-C_y B_x+A_x B_y+B_y C_x}\right)
   \left(B_x^2+B_y^2\right)}{\left(A_y B_x+C_y B_x-A_x B_y-B_y C_x\right){}^2} -\frac{\tan ^{-1}\left(\frac{-B_x^2+A_x B_x+C_x B_x+B_y \left(A_y-B_y+C_y\right)}{-A_y B_x-C_y B_x+A_x
   B_y+B_y C_x}\right) \left(B_x^2+B_y^2\right)}{\left(A_y B_x+C_y B_x-A_x B_y-B_y C_x\right){}^2} \nonumber \\
   & +\frac{\tan ^{-1}\left(\frac{A_x \left(C_x-B_x\right)+A_y \left(C_y-B_y\right)}{A_y
   \left(B_x-C_x\right)+A_x \left(C_y-B_y\right)}\right) \left(A_x^2+A_y^2\right)}{\left(A_y \left(B_x-C_x\right)+A_x \left(C_y-B_y\right)\right){}^2}-\frac{\tan
   ^{-1}\left(\frac{A_x^2+\left(C_x-B_x\right) A_x+A_y \left(A_y-B_y+C_y\right)}{A_y \left(B_x-C_x\right)+A_x \left(C_y-B_y\right)}\right) \left(A_x^2+A_y^2\right)}{\left(A_y
   \left(B_x-C_x\right)+A_x \left(C_y-B_y\right)\right){}^2} \Bigg) \label{coplanar}
\end{align}

\subsection*{Parallel cylinders}

\begin{align}
U =& \frac{K_\text{vol}}{2 \xi  \left(A_y C_x-A_x C_y\right)^3} \Bigg(A_x^2 \Bigg(\tan ^{-1}\left(\frac{A_x^2+C_x A_x+A_y \left(A_y+C_y\right)}{A_y C_x-A_x C_y}\right)\nonumber \\
&+\xi  \tan ^{-1}\left(\frac{-\xi  A_x^2+C_x A_x+A_y \left(C_y-\xi  A_y\right)}{A_x C_y-A_y
   C_x}\right)+\xi  \tan ^{-1}\left(\frac{(1-\xi ) A_x^2+C_x A_x+A_y \left(-\xi  A_y+A_y+C_y\right)}{A_y C_x-A_x C_y}\right)\nonumber \\
   &+ \tan ^{-1}\left(\frac{(1-\xi ) A_x^2+C_x A_x+A_y
   \left(-\xi  A_y+A_y+C_y\right)}{A_x C_y-A_y C_x}\right) \Bigg) \nonumber \\
   &+C_x A_x\Bigg(\tan ^{-1}\left(\frac{A_y C_x-A_x C_y}{A_x C_x+A_y C_y}\right)-\tan ^{-1}\left(\frac{A_y C_x-A_x
   C_y}{A_x^2+C_x A_x+A_y \left(A_y+C_y\right)}\right) \nonumber \\
   & +\tan ^{-1}\left(\frac{-\xi  A_x^2+C_x A_x+A_y \left(C_y-\xi  A_y\right)}{A_y C_x-A_x C_y}\right)-\tan ^{-1}\left(\frac{(1-\xi )
   A_x^2+C_x A_x+A_y \left(-\xi  A_y+A_y+C_y\right)}{A_y C_x-A_x C_y}\right) \Bigg)   \nonumber \\
   &+A_y \Bigg( A_y\Bigg(\tan ^{-1}\left(\frac{A_x^2+C_x A_x+A_y \left(A_y+C_y\right)}{A_y C_x-A_x
   C_y}\right)+\xi  \tan ^{-1}\left(\frac{-\xi  A_x^2+C_x A_x+A_y \left(C_y-\xi  A_y\right)}{A_x C_y-A_y C_x}\right) \nonumber \\ 
   &+\xi  \tan ^{-1}\left(\frac{(1-\xi ) A_x^2+C_x A_x+A_y \left(-\xi 
   A_y+A_y+C_y\right)}{A_y C_x-A_x C_y}\right) \nonumber \\
   &+\tan ^{-1}\left(\frac{(1-\xi ) A_x^2+C_x A_x+A_y \left(-\xi  A_y+A_y+C_y\right)}{A_x C_y-A_y C_x}\right)\Bigg)  \nonumber \\
   &+C_y\bigg(\tan
   ^{-1}\left(\frac{A_y C_x-A_x C_y}{A_x C_x+A_y C_y}\right)-\tan ^{-1}\left(\frac{A_y C_x-A_x C_y}{A_x^2+C_x A_x+A_y \left(A_y+C_y\right)}\right) \nonumber \\
   &+\tan ^{-1}\left(\frac{-\xi 
   A_x^2+C_x A_x+A_y \left(C_y-\xi  A_y\right)}{A_y C_x-A_x C_y}\right)-\tan ^{-1}\left(\frac{(1-\xi ) A_x^2+C_x A_x+A_y \left(-\xi  A_y+A_y+C_y\right)}{A_y C_x-A_x C_y}\right)\Bigg) \label{parallel}
   \Bigg)
\end{align}

\subsection*{Colinear cylinders}
\begin{align}
    U = \frac{K_\text{vol}}{6 \xi  \left(A_x^2+A_y^2\right){}^2} \Bigg(-\frac{1}{\zeta ^2}+\frac{1}{(\zeta -\xi )^2}-\frac{1}{(\zeta -\xi +1)^2}+\frac{1}{(\zeta +1)^2}\Bigg) \label{colinear}
\end{align}

\subsection*{The Gay-Berne potential}
The Gay-Berne potential is designed to generalize the familiar Lennard-Jones interaction to geometrically anisotropic ellipsoidal particles. Further generalizations to lower symmetry interactions have also been constructed, but we assume here a pair of identical radially symmetric ellipsoids repelling with a $1/r^4$ potential.  The formulas given here are adapted from Ref \citeonline{chen2016gpu}.  The interaction energy is written as 
\begin{equation}
    U(\mathbf{\hat{u}}_i, \mathbf{\hat{u}}_j, \mathbf{r}_{ij}) = 4 \epsilon_0 \epsilon(\mathbf{\hat{u}}_i, \mathbf{\hat{u}}_j, \mathbf{\hat{r}}_{ij})\left(\frac{\sigma_s}{r_{ij }-\sigma(\mathbf{\hat{u}}_i, \mathbf{\hat{u}}_j, \mathbf{\hat{r}}_{ij}) + \sigma_s }\right)^4. 
    \label{GBeq}
\end{equation}
Here $\mathbf{r}_{ij}$ points from the center of ellipsoid $i$ to the center of ellipsoid $j$, $\mathbf{u}_i$ points along the major axis of ellipsoid $i$ and likewise for $\mathbf{u}_j$, the caret hats indicate unit vectors, and $r_{ij}$ is the magnitude of $\mathbf{r}_{ij}$.   $\sigma_s$ represents the length of the minor ellipsoid axis (the `diameter'), and $\sigma_e$ represents the length of the major axis (the `length').  The prefactor $\epsilon_0$ sets the energy scale.  The shape function $\sigma(\mathbf{\hat{u}}_i, \mathbf{\hat{u}}_j, \mathbf{\hat{r}}_{ij})$ is 
\begin{equation}
    \sigma(\mathbf{\hat{u}_i}, \mathbf{\hat{u}}_j, \mathbf{\hat{r}}_{ij}) = \sigma_s \left(1 - \frac{\chi}{2} \left(\frac{(\mathbf{\hat{r}}_{ij} \cdot \mathbf{\hat{u}}_{i} + \mathbf{\hat{r}}_{ij} \cdot \mathbf{\hat{u}}_{j})^2}{1 + \chi \mathbf{\hat{u}}_{i} \cdot \mathbf{\hat{u}}_{j}} + \frac{(\mathbf{\hat{r}}_{ij} \cdot \mathbf{\hat{u}}_{i} - \mathbf{\hat{r}}_{ij} \cdot \mathbf{\hat{u}}_{j})^2}{1 - \chi \mathbf{\hat{u}}_{i} \cdot \mathbf{\hat{u}}_{j}} \right) \right)^{-1/2},
\end{equation}
where $\chi = \frac{\kappa^2 -1 }{\kappa^2 +1}$ and $\kappa = \frac{\sigma_e}{\sigma_s}$.  The interaction function $\epsilon(\mathbf{\hat{u}_i}, \mathbf{\hat{u}_j}, \mathbf{\hat{r}_{ij}})$ is 
\begin{equation}
    \epsilon(\mathbf{\hat{u}_i}, \mathbf{\hat{u}_j}, \mathbf{\hat{r}_{ij}}) = \left(1 - \chi^2 (\mathbf{\hat{u}}_{i} \cdot \mathbf{\hat{u}}_{j})^2  \right)^{-\frac{1}{2\nu}} \left(1 - \frac{\chi'}{2} \left(\frac{(\mathbf{\hat{r}}_{ij} \cdot \mathbf{\hat{u}}_{i} + \mathbf{\hat{r}}_{ij} \cdot \mathbf{\hat{u}}_{j})^2}{1 + \chi' \mathbf{\hat{u}}_{i} \cdot \mathbf{\hat{u}}_{j}} + \frac{(\mathbf{\hat{r}}_{ij} \cdot \mathbf{\hat{u}}_{i} - \mathbf{\hat{r}}_{ij} \cdot \mathbf{\hat{u}}_{j})^2}{1 - \chi' \mathbf{\hat{u}}_{i} \cdot \mathbf{\hat{u}}_{j}} \right) \right)^{\mu},
\end{equation}
where $\chi' = \frac{k'^{1/\mu}-1}{ k'^{1/\mu} + 1}$, $k' = \frac{\epsilon_s}{\epsilon_e}$, and $\epsilon_s$ and $\epsilon_e$ represent, respectively, the depth of the potential well for the side-to-side and end-to-end configurations of the two ellipsoids.  The free parameters of this energy are $\epsilon_0$, $\epsilon_s$, $\epsilon_e$, $\sigma_s$, $\sigma_e$, and the fitting exponents $\mu$ and $\nu$.  For the comparisons done in Figure \ref{GB}, we take $\sigma_s = 1$, $\epsilon_e = \epsilon_s = 1$, $\mu = 2$ and $\nu = 1$ (following Ref. \citeonline{chen2016gpu}), and $\epsilon_0$ is chosen so that $U = 1$ when the separation is $d^*$, as described in the main text.  $\sigma_e = \kappa \sigma_s$ is varied to test the effect of geometrical anisotropy.  

\bibliographystyle{unsrt}

\begin{thebibliography}{10}
	
	\bibitem{lopez2014vitro}
	Magdalena~Preciado L{\'o}pez, Florian Huber, Ilya Grigoriev, Michel~O
	Steinmetz, Anna Akhmanova, Marileen Dogterom, and Gijsje~H Koenderink.
	\newblock In vitro reconstitution of dynamic microtubules interacting with
	actin filament networks.
	\newblock In {\em Methods in Enzymology}, volume 540, pages 301--320. Elsevier,
	2014.
	
	\bibitem{brangwynne2008nonequilibrium}
	Clifford~P Brangwynne, Gijsje~H Koenderink, Frederick~C MacKintosh, and David~A
	Weitz.
	\newblock Nonequilibrium microtubule fluctuations in a model cytoskeleton.
	\newblock {\em Physical Review Letters}, 100(11):118104, 2008.
	
	\bibitem{doi2013soft}
	Masao Doi.
	\newblock {\em Soft Matter Physics}.
	\newblock Oxford University Press, 2013.
	
	\bibitem{rubinstein2003polymer}
	Michael Rubinstein, Ralph~H Colby, et~al.
	\newblock {\em Polymer Physics}, volume~23.
	\newblock Oxford university press New York, 2003.
	
	\bibitem{schiller2018mesoscopic}
	Ulf~D Schiller, Timm Kr{\"u}ger, and Oliver Henrich.
	\newblock Mesoscopic modelling and simulation of soft matter.
	\newblock {\em Soft Matter}, 14(1):9--26, 2018.
	
	\bibitem{gartner2019modeling}
	Thomas~E Gartner~III and Arthi Jayaraman.
	\newblock Modeling and simulations of polymers: A roadmap.
	\newblock {\em Macromolecules}, 52(3):755--786, 2019.
	
	\bibitem{howard2001mechanics}
	Jonathon Howard et~al.
	\newblock Mechanics of motor proteins and the cytoskeleton.
	\newblock 2001.
	
	\bibitem{li2020tensile}
	Xiaona Li, Qin Ni, Xiuxiu He, Jun Kong, Soon-Mi Lim, Garegin~A Papoian,
	Jerome~P Trzeciakowski, Andreea Trache, and Yi~Jiang.
	\newblock Tensile force-induced cytoskeletal remodeling: Mechanics before
	chemistry.
	\newblock {\em PLoS Computational Biology}, 16(6):e1007693, 2020.
	
	\bibitem{chandrasekaran2019remarkable}
	Aravind Chandrasekaran, Arpita Upadhyaya, and Garegin~A Papoian.
	\newblock Remarkable structural transformations of actin bundles are driven by
	their initial polarity, motor activity, crosslinking, and filament
	treadmilling.
	\newblock {\em PLoS Computational Biology}, 15(7):e1007156, 2019.
	
	\bibitem{floyd2019quantifying}
	Carlos Floyd, Garegin~A Papoian, and Christopher Jarzynski.
	\newblock Quantifying dissipation in actomyosin networks.
	\newblock {\em Interface Focus}, 9(3):20180078, 2019.
	
	\bibitem{freedman2019mechanical}
	Simon~L Freedman, Cristian Suarez, Jonathan~D Winkelman, David~R Kovar,
	Gregory~A Voth, Aaron~R Dinner, and Glen~M Hocky.
	\newblock Mechanical and kinetic factors drive sorting of f-actin cross-linkers
	on bundles.
	\newblock {\em Proceedings of the National Academy of Sciences},
	116(33):16192--16197, 2019.
	
	\bibitem{belmonte2017theory}
	Julio~M Belmonte, Maria Leptin, and Fran{\c{c}}ois N{\'e}d{\'e}lec.
	\newblock A theory that predicts behaviors of disordered cytoskeletal networks.
	\newblock {\em Molecular Systems Biology}, 13(9):941, 2017.
	
	\bibitem{freedman2017versatile}
	Simon~L Freedman, Shiladitya Banerjee, Glen~M Hocky, and Aaron~R Dinner.
	\newblock A versatile framework for simulating the dynamic mechanical structure
	of cytoskeletal networks.
	\newblock {\em Biophysical Journal}, 113(2):448--460, 2017.
	
	\bibitem{nedelec2007collective}
	Francois Nedelec and Dietrich Foethke.
	\newblock Collective langevin dynamics of flexible cytoskeletal fibers.
	\newblock {\em New Journal of Physics}, 9(11):427, 2007.
	
	\bibitem{kim2009computational}
	Taeyoon Kim, Wonmuk Hwang, Hyungsuk Lee, and Roger~D Kamm.
	\newblock Computational analysis of viscoelastic properties of crosslinked
	actin networks.
	\newblock {\em PLoS Computational Biology}, 5(7):e1000439, 2009.
	
	\bibitem{popov2016medyan}
	Konstantin Popov, James Komianos, and Garegin~A Papoian.
	\newblock Medyan: Mechanochemical simulations of contraction and polarity
	alignment in actomyosin networks.
	\newblock {\em PLoS Computational Biology}, 12(4):e1004877, 2016.
	
	\bibitem{broedersz2014modeling}
	Chase~P Broedersz and Fred~C MacKintosh.
	\newblock Modeling semiflexible polymer networks.
	\newblock {\em Reviews of Modern Physics}, 86(3):995, 2014.
	
	\bibitem{de1971reptation}
	Pierre-Giles de~Gennes.
	\newblock Reptation of a polymer chain in the presence of fixed obstacles.
	\newblock {\em The Journal of Chemical Physics}, 55(2):572--579, 1971.
	
	\bibitem{gotzelmann1998depletion}
	B~G{\"o}tzelmann, Robert Evans, and Siegfried Dietrich.
	\newblock Depletion forces in fluids.
	\newblock {\em Physical Review E}, 57(6):6785, 1998.
	
	\bibitem{chaikin1995principles}
	Paul~M Chaikin, Tom~C Lubensky, and Thomas~A Witten.
	\newblock {\em Principles of condensed matter physics}, volume~10.
	\newblock Cambridge university press Cambridge, 1995.
	
	\bibitem{phillips2012physical}
	Rob Phillips, Jane Kondev, Julie Theriot, and Hernan Garcia.
	\newblock {\em Physical biology of the cell}.
	\newblock Garland Science, 2012.
	
	\bibitem{rubinstein2012polyelectrolytes}
	Michael Rubinstein and Garegin~A Papoian.
	\newblock Polyelectrolytes in biology and soft matter.
	\newblock {\em Soft Matter}, 8(36):9265--9267, 2012.
	
	\bibitem{tadmor2002debye}
	Rafael Tadmor, Ernesto Hern{\'a}ndez-Zapata, Nianhuan Chen, Philip Pincus, and
	Jacob~N Israelachvili.
	\newblock Debye length and double-layer forces in polyelectrolyte solutions.
	\newblock {\em Macromolecules}, 35(6):2380--2388, 2002.
	
	\bibitem{angelini2006counterions}
	Thomas~E Angelini, Ramin Golestanian, Robert~H Coridan, John~C Butler,
	Alexandre Beraud, Michael Krisch, Harald Sinn, Kenneth~S Schweizer, and
	Gerard~CL Wong.
	\newblock Counterions between charged polymers exhibit liquid-like organization
	and dynamics.
	\newblock {\em Proceedings of the National Academy of Sciences},
	103(21):7962--7967, 2006.
	
	\bibitem{smith2011electrostatic}
	DA~Smith and DG~Stephenson.
	\newblock An electrostatic model with weak actin-myosin attachment resolves
	problems with the lattice stability of skeletal muscle.
	\newblock {\em Biophysical Journal}, 100(11):2688--2697, 2011.
	
	\bibitem{eghiaian2015structural}
	Fr{\'e}d{\'e}ric Eghiaian, Annafrancesca Rigato, and Simon Scheuring.
	\newblock Structural, mechanical, and dynamical variability of the actin cortex
	in living cells.
	\newblock {\em Biophysical Journal}, 108(6):1330--1340, 2015.
	
	\bibitem{sanders2005structure}
	Lori~K Sanders, Camilo Gu{\'a}queta, Thomas~E Angelini, Jae-Wook Lee, Scott~C
	Slimmer, Erik Luijten, and Gerard~CL Wong.
	\newblock Structure and stability of self-assembled actin-lysozyme complexes in
	salty water.
	\newblock {\em Physical Review Letters}, 95(10):108302, 2005.
	
	\bibitem{janmey1994mechanical}
	Paul~A Janmey, Soren Hvidt, J~K{\"a}s, Dietmar Lerche, Anthony Maggs, Erich
	Sackmann, Manfred Schliwa, and Thomas~P Stossel.
	\newblock The mechanical properties of actin gels. elastic modulus and filament
	motions.
	\newblock {\em Journal of Biological Chemistry}, 269(51):32503--32513, 1994.
	
	\bibitem{gay1981modification}
	JG~Gay and BJ~Berne.
	\newblock Modification of the overlap potential to mimic a linear site--site
	potential.
	\newblock {\em The Journal of Chemical Physics}, 74(6):3316--3319, 1981.
	
	\bibitem{berardi1998gay}
	Roberto Berardi, Carlo Fava, and Claudio Zannoni.
	\newblock A gay--berne potential for dissimilar biaxial particles.
	\newblock {\em Chemical Physics Letters}, 297(1-2):8--14, 1998.
	
	\bibitem{sirk2012enhanced}
	Timothy~W Sirk, Yelena~R Slizoberg, John~K Brennan, Martin Lisal, and Jan~W
	Andzelm.
	\newblock An enhanced entangled polymer model for dissipative particle
	dynamics.
	\newblock {\em The Journal of Chemical Physics}, 136(13):134903, 2012.
	
	\bibitem{plimpton1995fast}
	Steve Plimpton.
	\newblock Fast parallel algorithms for short-range molecular dynamics.
	\newblock {\em Journal of Computational Physics}, 117(1):1--19, 1995.
	
	\bibitem{mathematica}
	Wolfram~Research{,} Inc.
	\newblock Mathematica, {V}ersion 12.2.
	\newblock Champaign, IL, 2020.
	
	\bibitem{rich2018rule}
	Albert Rich, Patrick Scheibe, and Nasser~M Abbasi.
	\newblock Rule-based integration: An extensive system of symbolic integration
	rules.
	\newblock {\em Journal of Open Source Software}, 3(32):1073, 2018.
	
	\bibitem{rickman2019effects}
	Jamie Rickman, Fran{\c{c}}ois N{\'e}d{\'e}lec, and Thomas Surrey.
	\newblock Effects of spatial dimensionality and steric interactions on
	microtubule-motor self-organization.
	\newblock {\em Physical Biology}, 16(4):046004, 2019.
	
	\bibitem{letort2015geometrical}
	Ga{\"e}lle Letort, Antonio~Z Politi, Hajer Ennomani, Manuel Th{\'e}ry, Francois
	Nedelec, and Laurent Blanchoin.
	\newblock Geometrical and mechanical properties control actin filament
	organization.
	\newblock {\em PLoS Computational Biology}, 11(5):e1004245, 2015.
	
	\bibitem{bloomfield1996dna}
	Victor~A Bloomfield.
	\newblock Dna condensation.
	\newblock {\em Current Opinion in Structural Biology}, 6(3):334--341, 1996.
	
	\bibitem{leforestier2009structure}
	Am{\'e}lie Leforestier and Fran{\c{c}}oise Livolant.
	\newblock Structure of toroidal dna collapsed inside the phage capsid.
	\newblock {\em Proceedings of the National Academy of Sciences},
	106(23):9157--9162, 2009.
	
	\bibitem{Ni2021membrane}
	Haoran Ni and Garegin~A. Papoian.
	\newblock Membrane-medyan: Simulating deformable vesicles containing complex
	cytoskeletal networks.
	\newblock {\em bioRxiv}, 2021.
	
	\bibitem{chen2016gpu}
	Wenduo Chen, Youliang Zhu, Fengchao Cui, Lunyang Liu, Zhaoyan Sun, Jizhong
	Chen, and Yunqi Li.
	\newblock Gpu-accelerated molecular dynamics simulation to study liquid crystal
	phase transition using coarse-grained gay-berne anisotropic potential.
	\newblock {\em PLoS One}, 11(3):e0151704, 2016.
	
\end{thebibliography}

\end{document}